\documentclass{ws-rv9x6}
\usepackage{ws-rv-van}             
\begin{document}

\chapter{Neutrinoless double beta decay\label{bbdecay}}

\author{Petr Vogel}

\address{Kellogg Radiation Laboratory \\
Caltech, Pasadena, CA 91125, USA
\footnote{email: pxv@caltech.edu}
}

\begin{abstract}
The status of the search for neutrinoless double beta decay is reviewed.
The effort to reach the sensitivity needed to cover the effective
Majorana neutrino mass corresponding to the degenerate and
inverted mass hierarchy is described. Various issues concerning
the theory (and phenomenology) of the relation between the $0\nu\beta\beta$
decay rate and the absolute neutrino mass scale are discussed,
in particular the issue of  mechanism of the  $0\nu\beta\beta$ decay.
Finally, the relation between the neutrino magnetic moments
and the charge conjugation property (Dirac vs. Majorana) is described.
\end{abstract}

\body

\section{Introduction - fundamentals of $\beta\beta$ decay}\label{sec1.1}
In the recent past neutrino
oscillation experiments have convincingly
shown that neutrinos have a finite mass. However, in oscillation
experiments only the differences of squares of the neutrino masses, 
$\Delta m^2 \equiv | m_2^2 - m_1^2| $, can be measured, and the results
do not depend on the charge conjugation properties of neutrinos, i.e., whether
they are Dirac or Majorana fermions.
Nevertheless, a lower limit on the absolute value  of the neutrino mass scale, 
$m_{scale} =  \sqrt{|\Delta m^2|}$, has been
established in this way. Its existence, in turn, is causing a renaissance
of enthusiasm in the double beta decay community which is expected to reach
and even exceed, in
the next generation of experiments, the sensitivity corresponding to this mass scale. 
Below I review the current status of the double beta decay and the effort devoted
to reach the required sensitivity, as well as various issues 
in theory (or phenomenology) concerning the relation
of the $0\nu\beta\beta$ decay rate to the absolute neutrino mass scale and to
the general problem of the Lepton Number Violation (LNV). 

But before doing that I very briefly summarize
the achievements of the neutrino oscillation searches and the role that the search
for the neutrinoless double beta decay plays in the elucidation of the pattern
of neutrino masses and mixing. In these introductory remarks I use the 
established terminology, some of which will be defined only later in the text.  

There is a consensus that the measurement of atmospheric
neutrinos by the SuperKamiokande collaboration\cite{SKatm01}
can be only interpreted as a consequence
of the nearly maximum mixing between  $\nu_{\mu}$ and $\nu_{\tau}$ neutrinos,
with the corresponding mass squared difference  
$|\Delta m_{atm}^2| \sim  2.4\times10^{-3}{\rm eV}^2$.
This finding was confirmed  by the K2K experiment
\cite{K2K01} that uses accelerator  
$\nu_{\mu}$ beam pointing towards the SuperKamiokande detector
250 km away, as well as  by the very recent first result of the  MINOS 
experiment located at the Sudan mine in Minnesota, 735 km away from the 
Fermilab\cite{MINOS}. 
Several large long-baseline experiments are being built to further elucidate
this discovery, and determine the corresponding parameters more accurately. 

At the same time the ``solar neutrino puzzle", which has been with us for over thirty years
since the pioneering chlorine experiment of Davis\cite{chlorine}, also reached
the stage where the interpretation of the measurements in terms of  oscillations
between the $\nu_e$ and some combination of the active, 
i.e., $\nu_{\mu}$ and $\nu_{\tau}$ neutrinos, is inescapable. In particular, the
juxtaposition of the results of the
SNO experiment\cite{SNO01} and SuperKamiokande\cite{SKsol01},
together with the earlier solar neutrino flux determination in
the gallium experiments\cite{Gallex,Sage},
leads to that conclusion.
The value of the corresponding oscillation parameters,
however, remained uncertain, with several
``solutions" possible, although the so-called Large Mixing Angle (LMA) solution
with $\sin^2 2\theta_{sol} \sim 0.8$ and 
$\Delta m_{sol}^2 \sim 10^{-4}{\rm eV}^2$ was preferred. 
A decisive confirmation of the ``solar" oscillations was provided by
the nuclear reactor experiment KamLAND \cite{Kaml1,Kaml2} that
demonstrated that the flux of the reactor $\bar{\nu}_e$ is reduced
and its spectrum distorted at the average distance $\sim$ 180 km from 
nuclear reactors. 

\begin{figure}
\centerline{\psfig{file=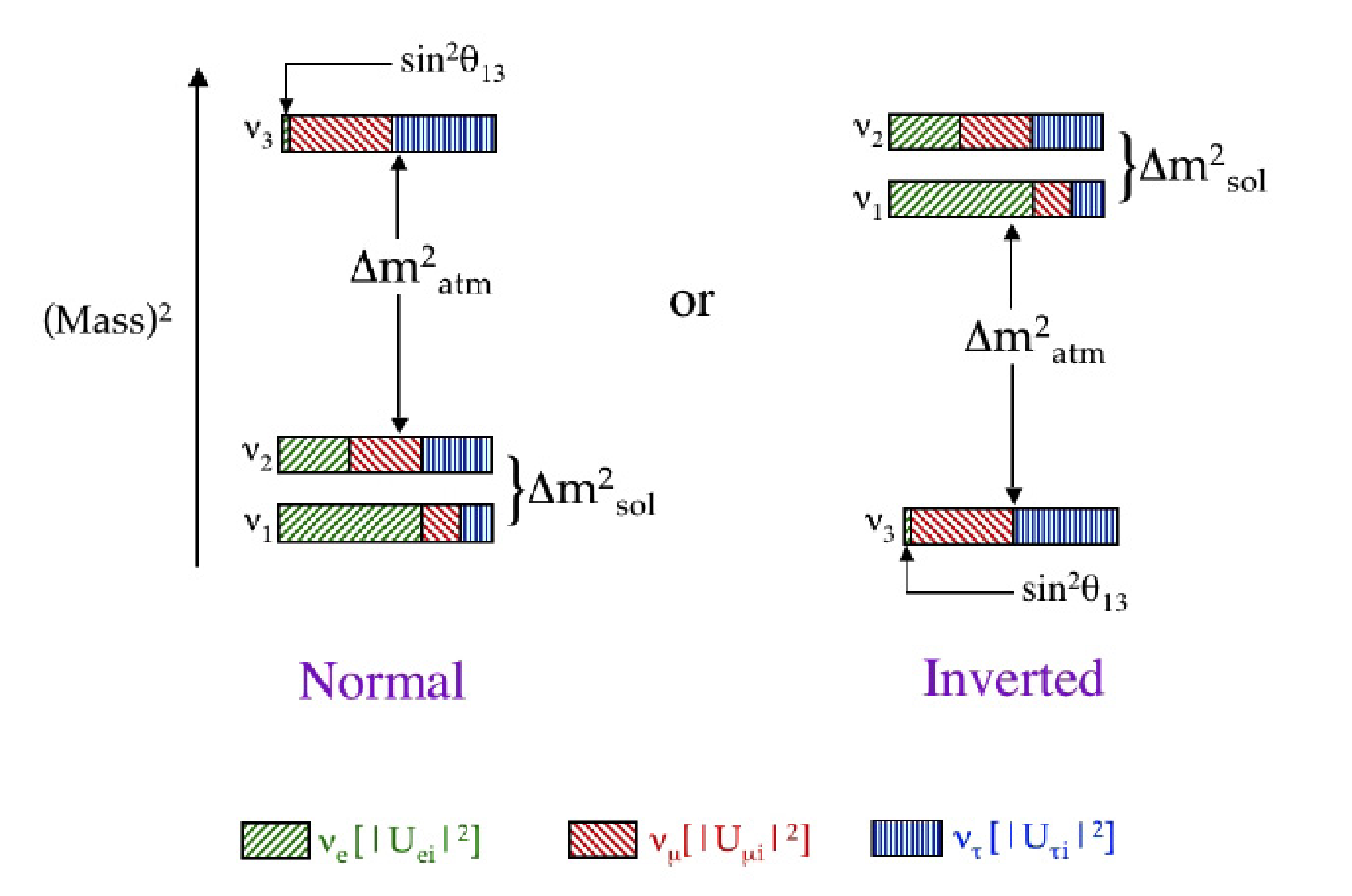,width=6.8cm}}
\caption{ Schematic illustration (mass intervals not to scale) of the decomposition
of the neutrino mass eigenstates $\nu_i$ in terms of the flavor eigenstates.
The two hierarchies cannot be, at this time, distinguished. The small admixture
of $\nu_e$ into $\nu_3$ is an upper limit. }
\label{fig_osc}
\end{figure}

The pattern of neutrino mixing is further simplified by the constraint due to
the Chooz and Palo Verde reactor neutrino experiments\cite{Chooz99,Palo01}
which lead to the
conclusion that the third mixing angle, $\theta_{13}$, is
small,  $\sin^2 2\theta_{13} \le 0.1$.  The two remaining possible 
neutrino mass patterns
are illustrated in Fig.\ref{fig_osc}.

Altogether, clearly a {\it lower}
limit for at least one of the neutrino masses, $\sqrt{\Delta m_{atm}^2} \simeq 0.05$ eV 
has been established.
However, the oscillation experiments cannot determine the absolute
magnitude of the masses and, in particular, cannot at this stage separate
two rather different scenarios, the hierarchical pattern of neutrino
masses in which $m \sim \sqrt{\Delta m^2}$ and the degenerate pattern
in which  $m \gg \sqrt{\Delta m^2}$. It is hoped that the 
search for the neutrinoless double beta
decay, reviewed here, will help in foreseeable future
 in determining, or at least narrowing down, the absolute neutrino mass
scale, and in deciding which of these two possibilities is applicable.

Moreover, the oscillation results do not tell us anything about the properties of neutrinos
under charge conjugation. While the charged leptons are Dirac particles, distinct from
their antiparticles, neutrinos may be the ultimate neutral particles, as
envisioned by Majorana, that are identical to their antiparticles. That fundamental
distinction becomes important only for massive particles and
becomes irrelevant in the massless limit.
Neutrinoless double beta decay proceeds only when neutrinos are massive 
Majorana particles, hence its observation would resolve the question.

Double beta decay ($\beta\beta$) is a nuclear transition 
$(Z,A) \rightarrow (Z+2,A)$ in which two neutrons
bound in a nucleus  are  simultaneously transformed into two protons plus two
electrons (and possibly other light neutral particles). This transition is
possible and potentially observable because 
nuclei with even $Z$ and $N$ are more bound than the odd-odd nuclei with
the same $A = N + Z$. 
Analogous transition of two protons into two neutrons
are also, in principle, possible
in several nuclei, but phase space considerations give preference to the former
mode.

\begin{figure}
\centerline{\psfig{file=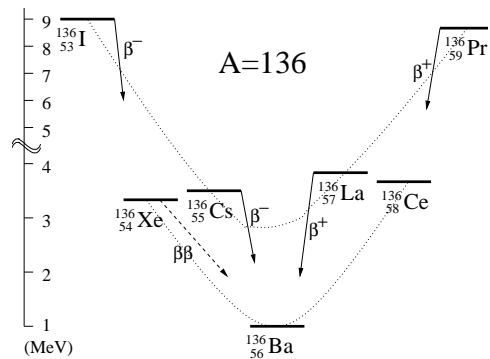,width=7.8cm}}
\caption{ Atomic masses of the isotopes with $A$ = 136. Nuclei $^{136}$Xe, $^{136}$Ba and
$^{136}$Ce are stable against the ordinary $\beta$ decay; hence they exist in nature. However,
energy conservation alone allows the transition
$^{136}$Xe $\rightarrow$ $^{136}$Ba + $2e^-$ (+ possibly
other neutral light particles) and the analogous decay of $^{136}$Ce with the positron emission. }
\label{fig_bb}
\end{figure}

An example is shown in Fig. \ref{fig_bb}. The situation shown there is not exceptional. 
There are eleven analogous cases (candidate nuclei) with the $Q$-value (i.e., the energy
available to leptons) in excess of 2 MeV.  

There are two basic modes of the $\beta\beta$ decay. In the two-neutrino mode ($2\nu\beta\beta$)
there are two $\bar{\nu}_e$ emitted together with the two $e^-$. Lepton number is conserved and
this mode is allowed in the standard model of electroweak interaction.
It has been repeatedly observed by now 
in a number of cases and proceeds with a typical half-life
of $\sim 10^{20}$years.
In contrast, in the neutrinoless
mode ($0\nu\beta\beta$) only the 2$e^-$ are emitted and nothing else. 
That mode clearly violates the law
of lepton number conservation and is forbidded in the standard model. Hence, its observation
would be a signal of a "new physics". 

The two modes of the $\beta\beta$ decay have some common and some distinct features.
The common features are:
\begin{itemize}
\item  The leptons carry essentially all available energy. The nuclear recoil is negligible,
$Q/Am_p \ll 1$.
\item The transition involves the $0^+$ ground state of the initial nucleus and
(in almost all cases) the $0^+$ ground state of the final nucleus. In few cases the
transition to an excited $0^+$ state in the final nucleus is energetically possible, but
suppressed by the smaller phase space available. (But the $2\nu\beta\beta$ decay to
the excited $0^+$ state has been observed in few cases.)
\item Both processes are of second order of weak interactions, $\sim G_F^4$, hence inherently
slow.  The phase space consideration alone (for the $2\nu\beta\beta$ mode $\sim Q^{11}$ and
for the $0\nu\beta\beta$ mode $\sim Q^5$) give preference to the $0\nu\beta\beta$ which is,
however, forbidden by the lepton number conservation.
\end{itemize}
The distinct features are:
\begin{itemize}
\item In the $2\nu\beta\beta$ mode the two neutrons undergoing the transition are uncorrelated
(but decay simultaneously) while in the $0\nu\beta\beta$ the two neutrons are correlated.
\item In the $2\nu\beta\beta$ mode the sum electron kinetic energy $T_1 + T_2$
spectrum is continuous and peaked below $Q/2$. As $T_1 + T_2 \rightarrow Q$ the
spectrum approaches zero approximately like $(\Delta E/Q)^6$.
 \item On the other hand, 
 in the $0\nu\beta\beta$ mode  $T_1 + T_2 = Q$ smeared only by the detector resolution.
 \end{itemize}
 
\begin{figure}
\centerline{\psfig{file=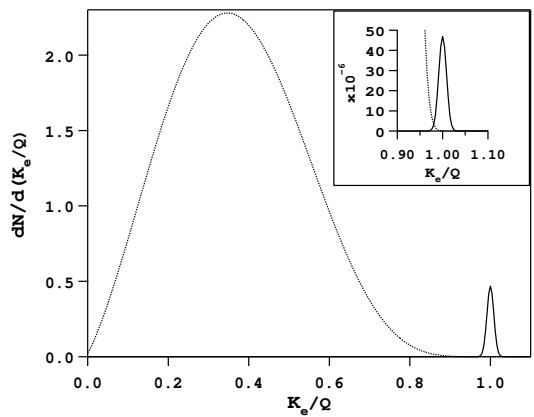,width=7.8cm}}
\caption{ Separating the $0\nu\beta\beta$ mode from the $2\nu\beta\beta$ by the shape of the
sum electron spectrum (kinetic energy $K_e$ of the two electrons), 
including the effect of the 2\% resolution smearing. The assumed $2\nu/0\nu$
rate ratio is $10^2$, and $10^6$ in the insert.}
\label{fig_2nu}
\end{figure}

These last features allow one to separate the two modes experimentally by observing the
sum electron spectrum with a good energy resolution, even if the corresponding decay rate
for the $0\nu\beta\beta$ mode is much smaller than for the $2\nu\beta\beta$ mode.
This is illustrated in Fig.\ref{fig_2nu} where the insert 
that includes the $0\nu$ peak and the $2\nu$ tail shows the situation for the rate
ratio of $1:10^6$ corresponding to the most sensitive current experiments.

Various aspects, both theoretical and experimental, 
of the $\beta\beta$ decay have been reviewed many times. Here I quote
just the more recent review articles\cite{FS98,Ver02,EV02,EE04}, earlier references
can be found there. 

\begin{figure}
\centerline{\psfig{file=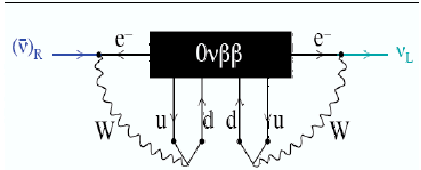,width=6.0cm}}
\caption{ By adding loops involving only standard weak interaction processes the
$0\nu\beta\beta$ decay amplitude (the black box)
implies the existence of the Majorana neutrino mass.  }
\label{fig_SV}
\end{figure}

In this introductory section let me make only few general remarks. 
The existence of the $0\nu\beta\beta$ decay
would mean that on the elementary particle level a six fermion 
lepton number violating amplitude
transforming two $u$ quarks into two $d$ quarks and two electrons
is nonvanishing.  As was first pointed out by Schechter and Valle\cite{SV82} 
more than twenty years ago,
this fact alone would guarantee that neutrinos are massive 
Majorana fermions (see Fig.\ref{fig_SV}). This qualitative statement (or theorem),
unfortunately, does not allow us to deduce the magnitude of the neutrino mass
once the rate of the $0\nu\beta\beta$ decay have been determined.

There is no indication at the present time that neutrinos have nonstandard interactions,
i.e. they seem to have only interactions carried by the $W$ and $Z$ bosons
that are contained in the Standard Electroweak Model.
All observed oscillation phenomena can be understood if one assumes
that that neutrinos interact exactly the way the Standard Model prescribes, but are massive
fermions forcing a generalization of the model. If we accept this, but in addition assume that
neutrinos are Majorana particles, we can in fact relate the   $0\nu\beta\beta$ decay  rate
to the quantity related to the absolute neutrino mass. With these caveats that relation 
can be expressed as 
\begin{equation}
\frac{1}{T_{1/2}^{0\nu}} = G^{0\nu}(Q,Z) |M^{0\nu}|^2 \langle m_{\beta\beta} \rangle^2 ~,
\label{eq_rate}
\end{equation}
where $G^{0\nu}(Q,Z)$ is a phase space factor that depends 
on the transition $Q$ value and through
the Coulomb effect on the emitted electrons on the nuclear charge $Z$ and that can be
easily and accurately calculated, $M^{0\nu}$ is the nuclear matrix element that can be
evaluated in principle, although with a considerable uncertainty, and finally the quantity
$\langle m_{\beta\beta} \rangle$ is the effective neutrino Majorana mass, representing
the important particle physics ingredient of the process.

In turn, the effective mass $\langle m_{\beta\beta} \rangle$ is related 
to the mixing angles $\theta_{ij}$
that are determined or constrained by the oscillation experiments, to the absolute neutrino
masses $m_i$ of the mass eigenstates $\nu_i$ and to the 
as of now totally unknown additional
parameters as fundamental as the mixing angles $\theta_{ij}$, 
the so-called Majorana phase $\alpha(i)$,
\begin{equation}
 \langle m_{\beta\beta} \rangle = | \Sigma_i |U_{ei}|^2 e^{i \alpha (i)} m_i | ~.
 \label{eq_mbb}
 \end{equation}
Here $U_{ei}$ are the matrix elements of the first row of the neutrino mixing matrix.

\begin{figure}
\centerline{\psfig{file=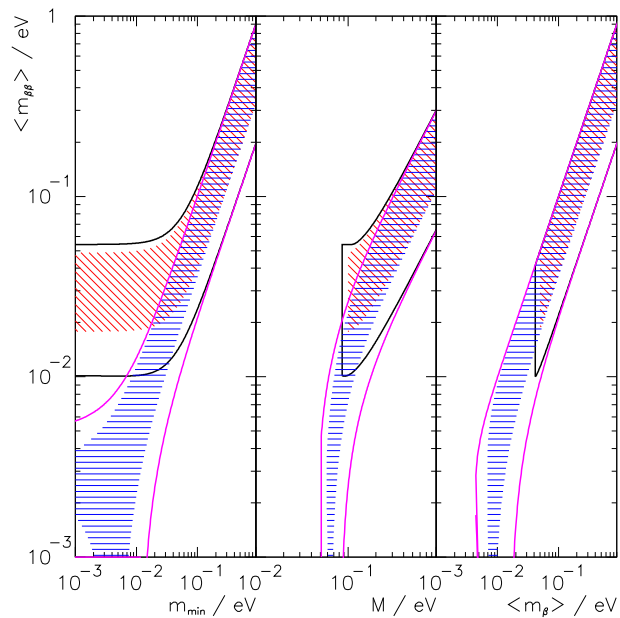,width=9.0cm}}
\caption{The left panel shows the dependence of $\langle m_{\beta\beta} \rangle$ on the  
mass of the lightest neutrino $m_{min}$, the middle one shows the relation between
 $\langle m_{\beta\beta} \rangle$ and the sum of   neutrino masses $M = \Sigma m_i$
 determined or constrained by the ``observational cosmology", and the right
 one depicts the relation between $\langle m_{\beta\beta} \rangle$ and the effective mass 
 $\langle m_{\beta} \rangle$ determined or 
constrained by the ordinary $\beta$ decay. In all panels the
 width of the hatched area is due to the unknown Majorana phases and therefore irreducible. 
 The solid lines indicate the allowed regions 
by taking into account the current uncertainties in the
 oscillation parameters; they will shrink as the accuracy improves. The two sets of curves  
 correspond to the normal and inverted hierarchies, they merge above about 
 $\langle m_{\beta\beta} \rangle \ge$ 0.1 eV, where the degenerate mass pattern begins.  }
\label{fig_5}
\end{figure}

It is straightforward to use the eq.(\ref{eq_mbb}) and the known 
neutrino oscillation results in order to relate $\langle m_{\beta\beta} \rangle$
to other neutrino mass dependent quantities. This is illustrated in Fig.\ref{fig_5}.
Traditionally such plot is made as in the left panel. However, the lightest
neutrino mass $m_{min}$ is not an observable quantity. For that reason
the other two panels show the relation of $\langle m_{\beta\beta} \rangle$
to the sum of the neutrino masses $M = \Sigma m_i$ and also to $\langle m_{\beta} \rangle$
that represents the parameter that can be determined or constrained in ordinary
$\beta$ decay,
\begin{equation}
\langle m_{\beta} \rangle^2 = \Sigma_i |U_{ei}|^2 m_i^2 ~.
\end{equation}

Several remarks are in order. First, the observation of the $0\nu\beta\beta$ decay 
and determination of $\langle m_{\beta\beta} \rangle$, even when combined
with the knowledge of $M$ and/or $\langle m_{\beta} \rangle$ does not allow,
in general, to distinguish between the normal and inverted mass orderings. This is
a consequence of the fact that the Majorana phases are unknown. In regions in
Fig. \ref{fig_5} where the two hatched bands overlap it is clear that two solutions
with the same  $\langle m_{\beta\beta} \rangle$ and the same $M$ 
(or the same $\langle m_{\beta} \rangle$) exist and cannot be distinguished.

On the other hand, obviously, if one can determine that
$\langle m_{\beta\beta} \rangle \ge$ 0.1 eV we would conclude that the 
mass pattern is degenerate. And in the so far hypothetical case
that one could show that $\langle m_{\beta\beta} \rangle \le $ 
0.01 - 0.02 eV but nonvanishing
nevertheless the normal hierarchy would be established\footnote{In that case
also the $\langle m_{\beta} \rangle$ in the right panel would not represent the quatity
directly related to the ordinary $\beta$ decay. There are no 
realistic ideas, however, how to reach 
the corresponding sensitivity in ordinary $\beta$ decay at this time.}.

It is worthwhile noting that if the inverted mass ordering is realized in nature,
(and neutrinos are Majorana particles) the quantity $\langle m_{\beta\beta} \rangle$
is constrained from below by $\sim$ 0.01 eV. This value is within reach of the
next generation of experiments. Also, in principle, in the
case of the normal hierarchy while all neutrinos could be massive
Majorana particles it is still possible that 
$\langle m_{\beta\beta} \rangle$ = 0. Such a situation, however, requires ``fine tuning"
or reflects a symmetry of some kind.

Let us remark that the $0\nu\beta\beta$ decay is not the only LNV process for which 
important experimental constraints exist. Examples of the other analogous
processes are
\begin{eqnarray}
&& \mu^- + (Z,A)  \rightarrow  e^+ + (Z-2,A); {~\rm exp.~ branching~ ratio} \le 10^{-12} ~,
\nonumber \\
&&  K^+  \rightarrow   \mu^+ \mu^+ \pi^-;  {~\rm exp. ~branching ~ratio} \le 3 \times 10^{-9} ~,
  \nonumber \\
&&  \bar{\nu}_e {\rm ~emission~ from~ the~ Sun};  {~\rm exp.~ branching~ ratio} \le 10^{-4} ~.
\end{eqnarray}  
However, detailed analysis suggests that the study of the $0\nu\beta\beta$ decay is by far the
most sensitive test of LNV. In simple terms, this is caused by the amount of tries one can make.
A 100 kg $0\nu\beta\beta$ decay source contains $\sim 10^{27}$ nuclei. This can be contrasted
with the possibilities of first producing muons or kaons, and then searching for the unusual
decay channels. The Fermilab accelerators, for example, produce 
``a few'' $\times 10^{20}$ protons on target per year
in their beams and thus correspondingly smaller numbers of muons or kaons.

\section{Mechanism of the $0\nu\beta\beta$ decay}

It has been recognized long time ago that the relation between the $0\nu\beta\beta$ decay
rate and the effective Majorana mass $\langle m_{\beta\beta} \rangle$ is to some extent
problematic. The assumption leading to the 
eq.(\ref{eq_rate}) is rather conservative, namely  that there is an exchange
of a virtual light, but massive, 
Majorana neutrino between the two nucleons undergoing the transition,
and that these neutrinos interact by the standard left-handed weak currents. 
However, that is not
the only possible mechanism. LNV interactions involving so far unobserved heavy ($\sim$ TeV)
particles
can lead to a comparable $0\nu\beta\beta$ decay rate. Thus, in the absence of additional
information about the mechanism responsible for the $0\nu\beta\beta$ decay,
one could not unambiguously infer  $\langle m_{\beta\beta} \rangle$  
from the $0\nu\beta\beta$ decay rate.

In general $0\nu\beta\beta$ decay can be generated by (i) light massive Majorana 
neutrino exchange or (ii) heavy particle exchange (see, e.g. Refs.\cite{heavy,Pre03}),
resulting from LNV dynamics at some scale $\Lambda$ above the electroweak one.
The relative size of heavy ($A_H$) versus light 
particle ($A_L$) exchange contributions to the decay amplitude 
can be crudely estimated as follows~\cite{Mohapatra:1998ye}: 
\begin{equation}
A_L \sim G_F^2  \frac{\langle m_{\beta \beta} \rangle}{\langle k^2 \rangle}  ,~ 
 A_H \sim G_F^2  \frac{M_W^4}{\Lambda^5}  ,~
\frac{A_H}{A_L} \sim \frac{M_W^4 \langle k^2 \rangle } 
{\Lambda^5  \langle m_{\beta \beta} \rangle }  \ , 
\label{eq_estimate}
\end{equation}
where $\langle m_{\beta \beta} \rangle$ is the effective neutrino
Majorana mass, 
$\langle k^2 \rangle \sim ( 50 \ {\rm MeV} )^2 $ is the
typical light neutrino virtuality, and $\Lambda$ is the heavy
scale relevant to the LNV dynamics. 
Therefore,  $A_H/A_L \sim O(1)$ for  $\langle m_{\beta \beta} \rangle \sim 0.1-0.5$ 
eV and $\Lambda \sim 1$ TeV, and  thus the LNV dynamics at the TeV
scale leads to similar $0 \nu \beta \beta$ decay rate as the
exchange of light Majorana neutrinos with the effective mass 
$\langle m_{\beta \beta} \rangle \sim 0.1-0.5$ eV. 

Obviously, the lifetime measurement by itself
does not provide the means for determining the underlying mechanism.
The spin-flip and non-flip exchange can be, in principle,
distinguished by the measurement of the single-electron spectra or
polarization (see e.g., \cite{Doi}).  However, in most cases the
mechanism of light Majorana neutrino exchange, and of
heavy particle exchange cannot be separated by the observation
of the emitted electrons. Thus one must look for other phenomenological
consequences of the different mechanisms other than observables 
directly associated with $0\nu\beta\beta$. Here I discuss the
suggestion\cite{LNVus} that under natural assumptions the presence of low scale
LNV interactions also affects muon lepton flavor violating (LFV)
processes, and in  particular enhances the $\mu \to e$ conversion 
compared to the $\mu \to e \gamma$ decay.

The discussion is
concerned mainly with the branching ratios $B_{\mu \rightarrow e \gamma} = \Gamma
(\mu \rightarrow e \gamma)/ \Gamma_\mu^{(0)}$ and $B_{\mu \to e} =
\Gamma_{\rm conv}/\Gamma_{\rm capt} $, where $\mu \to e \gamma$ is
normalized to the standard muon decay rate $\Gamma_\mu^{(0)} = (G_F^2
m_\mu^5)/(192 \pi^3)$, while $\mu \to e$ conversion is normalized to
the corresponding capture rate $\Gamma_{\rm capt}$. The main diagnostic tool in the
analysis is the ratio 
\begin{equation}
{\cal R} = B_{\mu \to e}/B_{\mu \rightarrow e \gamma} ~,
\end{equation}
and the relevance of our observation relies on the potential
for LFV discovery in the forthcoming experiments  MEG~\cite{MEG}
($\mu \to e \gamma$) and MECO~\cite{MECO} 
($\mu \to e$ conversion)\footnote{Even though MECO 
experiment was recently cancelled, proposals
for experiments with similar sensitivity exist elsewhere.}
that plan to improve the current limits by several orders of
magnitude.

It is useful to formulate the problem in terms of effective low energy
interactions obtained after integrating out the heavy degrees of
freedom that induce LNV and LFV dynamics. If the scales for both 
LNV and LFV are well above the weak scale, then one would not 
expect to observe any signal in the forthcoming LFV experiments, nor would 
the effects of heavy particle exchange enter $0\nu\beta\beta$ 
at an appreciable level. In this case, the only origin of a signal in 
$0\nu\beta\beta$ at the level of prospective experimental sensitivity 
would be the exchange of a light Majorana neutrino, leading to eq.(\ref{eq_rate}),
and allowing one to extract  $\langle m_{\beta \beta} \rangle$ from the decay rate.

In general, however, the two scales may be distinct, as in
SUSY-GUT~\cite{Barbieri:1995tw} or SUSY see-saw~\cite{Borzumati:1986qx} models. 
In these scenarios, both the Majorana neutrino mass as well as LFV effects are 
generated at the GUT scale.
The effects of heavy Majorana neutrino exchange in $0\nu\beta\beta$ are, thus, 
highly suppressed. In contrast,  the effects of GUT-scale LFV are transmitted 
to the TeV-scale by a soft SUSY-breaking sector without mass suppression 
via renormalization group running of the high-scale LFV couplings. 
Consequently, such scenarios could lead to observable effects 
in the upcoming LFV experiments but with an ${\cal O}(\alpha)$ 
suppression of the branching ratio 
$B_{\mu\to e}$ relative to $B_{\mu\to e\gamma}$ 
due to the exchange of a virtual photon in the conversion process 
rather than the emission of a real one,
thus ${\cal R} \sim 10^{-(2-3)}$ in this case.

The case where the
scales of LNV and LFV are both relatively low ($\sim$ TeV)
is more subtle.
This is the scenario which might lead to observable signals
in LFV searches and at the same time generate ambiguities in
interpreting a  positive signal in $0 \nu \beta \beta$.
Therefore, this is the case where one needs to develop some
discriminating criteria.

Denoting the new physics scale by $\Lambda$, one has a LNV
effective lagrangian of the form
\begin{equation}
{\cal L}_{0 \nu \beta \beta} = \displaystyle\sum_i \
\frac{\tilde{c}_i}{\Lambda^5}  \  \tilde{O}_i
\qquad  \tilde{O}_{i} =  \bar{q} \Gamma_1 q \,  \
\bar{q} \Gamma_2 q \,   \bar{e} \Gamma_3 e^c   \ ,
\label{eq:lag1}
\end{equation}
where we have suppressed the flavor and Dirac structures
(a complete list of the dimension nine operators
$\tilde{O}_i$ can be found in Ref.~\cite{Pre03}).

For the LFV interactions, one has
\begin{equation}
{\cal L}_{\rm LFV} = \displaystyle\sum_i \
\frac{c_i}{\Lambda^2}  \  O_i  \  ,
\label{eq:lag2}
\end{equation}
and a complete operator basis can be found in
Refs.\cite{Raidal:1997hq,Kitano:2002mt}.  The LFV operators relevant to
our analysis are of the following type (along with their analogues
with $L \leftrightarrow R$):
\begin{eqnarray}
O_{\sigma L} & = &   \displaystyle\frac{e}{(4 \pi)^2}
 \overline{\ell_{iL}} \, \sigma_{\mu \nu} i
/ \hspace{-0.23cm}D \, \ell_{jL}  \  F^{\mu \nu}  + {\rm h.c.}
\nonumber \\
O_{\ell L} & = &   \overline{\ell_{iL}} \, \ell^c_{jL} \
\overline{\ell^c_{kL}} \, \ell_{mL}
\nonumber \\
O_{\ell q} & = &   \overline{\ell_{i}} \Gamma_\ell \ell_{j} \
\overline{q} \Gamma_q  q  \  .
\end{eqnarray}

Operators of the type $O_{\sigma}$ are typically generated at one-loop level,
hence our choice to explicitly display the loop factor
$1/(4 \pi)^2$.   On the
other hand, in a large class of models, operators of the type $O_{\ell}$ or
$O_{\ell q}$ are generated by tree level exchange of heavy degrees of
freedom. With the above choices,
all non-zero $c_i$ and $\tilde{c}_i$ are nominally of the same size,
typically the product of two Yukawa-like couplings or gauge couplings
(times flavor mixing matrices). 

\begin{figure}
\centerline{\psfig{file=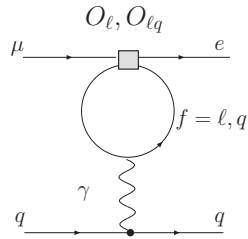,width=3.5cm}}
\caption{Loop contributions to $\mu \to e$ conversion through insertion of 
operators $O_{\ell}$ or $O_{\ell q}$, generating the large logarithm.}
\label{fig_6}
\end{figure}

With the notation established above, the ratio ${\cal R}$
of the branching ratios $\mu \to e$ to $\mu \to e + \gamma$  can be 
written schematically as follows (neglecting flavor indices in the
effective couplings and the term with $L \leftrightarrow R$):
\begin{eqnarray}  
{\cal R} &=& 
\displaystyle\frac{\Phi}{48 \pi^2} \,
\Big| \lambda_1  \, e^2  c_{\sigma L}  + e^2 \left( 
\lambda_2   c_{\ell L} + \lambda_3 c_{\ell q} \right)   
\log \displaystyle\frac{\Lambda^2}{m_\mu^2} 
\nonumber \\
&+&   \lambda_4 (4 \pi)^2  c_{\ell q} \ + \dots 
\Big|^2 / 
\left[ e^2 \left( |c_{\sigma L}|^2  + |c_{\sigma R}|^2 \right) \right] \, .
\label{eq:main1}  
\end{eqnarray}
In the above formula $\lambda_{1,2,3,4}$ are numerical factors of
$O(1)$, while the overall factor $\frac{\Phi}{48 \pi^2}$ 
arises from phase space and overlap
integrals of electron and muon wavefunctions in the nuclear field. For
light nuclei $\Phi = (Z F_p^2)/(g_V^2 + 3 g_A^2) \sim O(1)$ ($g_{V,A}$
are the vector and axial nucleon form factors at zero momentum
transfer, while $F_p$ is the nuclear form factor at
$q^2 = -m_\mu^2$~\cite{Kitano:2002mt}). The dots indicate 
subleading terms, not
relevant for our discussion, such as loop-induced
contributions to $c_{\ell}$ and $c_{\ell q}$ that are analytic in
external masses and momenta.  In contrast the 
logarithmically-enhanced loop contribution given by the second term in
the numerator of ${\cal R}$ plays an essential role. This term arises 
whenever the operators $O_{\ell L,R}$ and/or $O_{\ell q}$ appear at
tree-level in the effective theory and generate one-loop
renormalization of $O_{\ell q}$~\cite{Raidal:1997hq} (see
Fig.~\ref{fig_6}).

The ingredients in eq.~(\ref{eq:main1}) lead to several observations:
(i) In absence of tree-level $c_{\ell L }$ and $c_{\ell q}$, one
obtains ${\cal R} \sim (\Phi \, \lambda_1^2 \, \alpha)/(12 \pi) \sim
10^{-3}-10^{-2}$, due to gauge coupling and phase space
suppression. 
(ii) When present, the logarithmically enhanced contributions
compensate for the gauge coupling and phase space suppression, leading
to ${\cal R} \sim O(1)$. 
(iii) If present, the tree-level coupling $c_{\ell q}$ dominates the
$\mu \to e$ rate leading to ${\cal R} \gg 1$. 

Thus, we can formulate our main conclusions regarding the discriminating
power of the ratio ${\cal R}$:
\begin{enumerate} 
\item 
Observation of both the LFV muon processes
$\mu \to e$ and $\mu \to e \gamma$ with relative ratio ${\cal R} \sim
10^{-2}$ implies, under generic conditions, that $\Gamma_{0 \nu \beta
\beta} \sim \langle m_{\beta \beta} \rangle^2$. Hence the relation
of the $0\nu\beta\beta$ lifetime to the absolute neutrino mass scale
is straightforward.
\item 
On the other hand, observation of LFV muon processes with
relative ratio ${\cal R} \gg 10^{-2}$ could signal non-trivial LNV
dynamics at the TeV scale, whose effect on $0 \nu \beta \beta$ has to
be analyzed on a case by case basis. Therefore, in this scenario no
definite conclusion can be drawn based on LFV rates.
\item
Non-observation of LFV in muon processes in forthcoming 
experiments would imply either that the scale of non-trivial LFV and
LNV is  above a few TeV, and thus $\Gamma_{0 \nu \beta
\beta} \sim \langle m_{\beta \beta} \rangle^2$, or that any TeV-scale LNV is
approximately flavor diagonal.
\end{enumerate}

The above statements are illustrated using two explicit cases\cite{LNVus}: the
minimal supersymmetric standard model (MSSM) with R-parity violation
(RPV-SUSY) and the Left-Right Symmetric Model (LRSM).

\begin{figure}
\centerline{\psfig{file=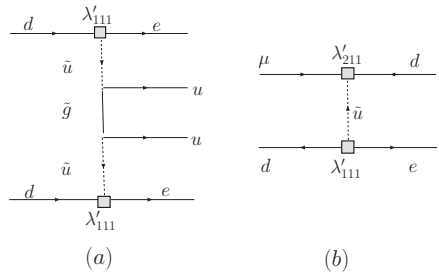,width=6.0cm}}
\caption{Gluino exchange contribution to $0 \nu \beta \beta$ $(a)$, 
and typical tree-level contribution to $O_{\ell q}$ $(b)$ in RPV SUSY.}
\label{fig_7}
\end{figure}

{\em RPV SUSY ---}  If one does not impose R-parity conservation [$R=
(-1)^{3 (B-L) + 2 s}$], the MSSM superpotential includes, in addition
to the standard Yukawa terms, lepton and baryon number violating
interactions, compactly written as (see e.g.,~\cite{Dreiner:1997uz})  
\begin{eqnarray}
W_{RPV} &=& \lambda_{ijk} L_i L_j E_k^c + 
\lambda_{ijk}' L_i Q_j D_k^c + 
\lambda_{ijk}''  U_i^c D_j^c D_k^c 
\nonumber \\
& & +  \mu_{i}' L_i  H_u   \ , 
\end{eqnarray}
where $L$ and $Q$ represent lepton and quark doublet superfields,
while $E^c$, $U^c$, $D^c$ are lepton and quark singlet superfields.
The simultaneous presence of $\lambda'$ and $\lambda ''$ couplings
would lead to an unacceptably large proton decay rate (for SUSY mass
scale $\Lambda_{SUSY} \sim$ TeV), so we focus on the case of
$\lambda'' = 0$ and set $\mu'=0$ without loss of
generality. In such case, lepton number is
violated by the remaining terms in $W_{RPV}$, leading to short
distance contributions to $0 \nu \beta \beta$
[e.g., Fig.\ref{fig_7}(a)], with typical
coefficients [cf. eq.~(\ref{eq:lag1})]
\begin{equation} 
\frac{\tilde{c_i}}{\Lambda^5} \sim 
\frac{\pi
\alpha_s}{m_{\tilde{g}}} \frac{\lambda_{111}'^2}{m_{\tilde{f}}^4} \, ; \, 
\frac{\pi
\alpha_2}{m_\chi} \frac{\lambda_{111}'^2}{m_{\tilde{f}}^4} \ ,  
\end{equation}
where $\alpha_s, \alpha_2$ represent the strong and weak gauge
coupling constants, respectively.  The RPV interactions also lead to
lepton number conserving but lepton flavor violating operators [e.g.
 Fig.~\ref{fig_7}(b)], with coefficients
[cf. eq.~(\ref{eq:lag2})]
\begin{eqnarray}
\frac{c_{\ell}}{\Lambda^2} &\sim &
\frac{ \lambda_{i11} \lambda_{i21}^*}{m_{\tilde{\nu}_i}^2} , 
\frac{ \lambda_{i11}^* \lambda_{i12}}{m_{\tilde{\nu}_i}^2}  \ , 
\nonumber \\
\frac{c_{\ell q}}{\Lambda^2} & \sim & 
\frac{ \lambda_{11i}'^* \lambda_{21i}'}{m_{\tilde{d}_i}^2}, 
\frac{ \lambda_{1i1}'^* \lambda_{2i1}'}{m_{\tilde{u}_i}^2} \ , 
\nonumber \\
\frac{c_\sigma}{\Lambda^2} 
& \sim &  \frac{ \lambda \lambda^* }{m_{\tilde{\ell}}^2}, 
\frac{\lambda'  \lambda'^* }{m_{\tilde{q}}^2}  \ ,
\end{eqnarray}
where the flavor combinations contributing to $c_{\sigma}$ can be
found in Ref.~\cite{deGouvea:2000cf}.  Hence, for generic flavor
structure of the couplings $\lambda$ and $\lambda'$ 
the underlying LNV dynamics generate both short distance
contributions to $0 \nu \beta \beta$ and LFV contributions that lead
to ${\cal R} \gg 10^{-2}$.  

Existing limits on rare processes strongly constrain combinations of
RPV couplings, assuming $\Lambda_{SUSY}$ is 
between a few hundred GeV and $\sim$ 1 TeV.  Non-observation of LFV at future
experiments MEG and MECO could be attributed either to a larger 
$\Lambda_{SUSY}$ ($>$ few TeV) or to suppression of couplings that
involve mixing among first and second generations.  In the former
scenario, the short distance contribution to $0\nu \beta 
\beta$ does not compete with the long distance one
[see eq.~(\ref{eq_estimate})], so that $\Gamma_{0 \nu \beta \beta}
\sim \langle m_{\beta \beta} \rangle^2$.  On the other hand, there is an exception to this
"diagnostic tool". If the $\lambda$ and
$\lambda'$ matrices are nearly flavor diagonal, the exchange of
superpartners may still make non-negligible contributions to $0\nu
\beta \beta$ without enhancing the ratio ${\cal R}$ .

{\em LRSM ---} The LRSM provides a natural scenario for introducing
non-sterile, right-handed neutrinos and Majorana
masses~\cite{Mohapatra:1979ia}.  The corresponding electroweak gauge
group $SU(2)_L \times SU(2)_R \times U(1)_{B-L}$, breaks down to
$SU(2)_L \times U(1)_Y$ at the scale $\Lambda \ge {\cal O}({\rm
TeV})$.  The symmetry breaking is implemented through an extended
Higgs sector, containing a bi-doublet $\Phi$ and two triplets
$\Delta_{L,R}$, whose leptonic couplings generate both Majorana
neutrino masses and LFV involving charged leptons:

\begin{eqnarray}
{\cal L}_Y^{\rm lept} &=& - \  
\overline{L_L}\, ^{i} \, 
\left( 
 y_D^{ij}  \,  \Phi \ + \   \tilde{y}_D^{ij} \,  \tilde{\Phi} 
\right) \, 
L_{R}^j  \\
&-&  
\overline{(L_{L})^c}\, ^i \    y_M^{ij} \, \tilde{\Delta}_L \  L_{L}^j
\ - \ 
 \overline{(L_{R})^c}\, ^i \  y_M^{ij}  \, \tilde{\Delta}_R  \ L_{R}^j \ . 
\nonumber 
\end{eqnarray}
Here $ \tilde{\Phi} = \sigma_2 \Phi^* \sigma_2$,
$\tilde{\Delta}_{L,R} = i \sigma_2 \Delta_{L,R}$, and leptons belong 
to two isospin doublets $L_{L,R}^i = (\nu_{L,R}^i,
\ell_{L,R}^i)$.  The gauge symmetry is broken through the VEVs
$\langle \Delta^0_R \rangle = v_R$, $\langle \Delta^0_L \rangle = 0$,
$ \langle \Phi \rangle = {\rm diag}(\kappa_1, \kappa_2) $. 
After diagonalization of the lepton mass matrices, LFV arises from
both non-diagonal gauge interactions and the Higgs Yukawa
couplings. In particular, the $\Delta_{L,R}$-lepton interactions are
not suppressed by lepton masses and have the structure ${\cal L} \sim
\Delta_{L,R}^{++} \, \overline{\ell_i^c} \, h_{ij} \, (1 \pm \gamma_5)
\ell_j + {\rm h.c.}$. The couplings $h_{ij}$ are in general
non-diagonal and related to the heavy neutrino mixing
matrix~\cite{Cirigliano:2004mv}.

\begin{figure}
\centerline{\psfig{file=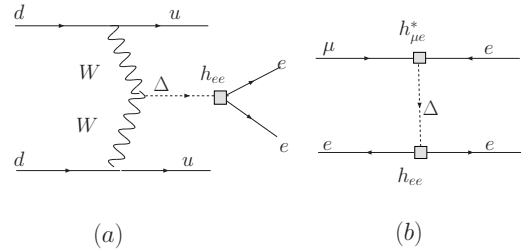,width=7.0cm}}
\caption{Typical doubly charged Higgs contribution to $0 \nu \beta \beta$ $(a)$ 
and to $O_{\ell}$ $(b)$ in the LRSM.}
\label{fig_8}
\end{figure}

Short distance contributions to $0\nu \beta \beta$ arise from the
exchange of both heavy $\nu$s and $\Delta_{L,R}$ 
(Fig.\ref{fig_8}a), with
\begin{equation}
\frac{\tilde{c}_i}{\Lambda^5} \sim  
\frac{g_2^4}{M_{W_R}^4} \frac{1}{M_{\nu_R}} 
\,  ; \, \frac{g_2^3}{M_{W_R}^3} \frac{h_{ee}}{M_\Delta^2} \ , 
\end{equation}
where $g_2$ is the weak gauge coupling. LFV operators are also generated 
through non-diagonal gauge and Higgs vertices, with~\cite{Cirigliano:2004mv}
(Fig.\ref{fig_8}b)
\begin{equation}
\frac{c_{\ell}}{\Lambda^2} \sim \frac{h_{\mu i} h_{ie}^*}{m_{\Delta}^2} \qquad 
\frac{c_{\sigma}}{\Lambda^2} \sim 
\frac{(h^\dagger h)_{e \mu}}{M_{W_R}^2}  \quad   i=e, \mu, \tau \ . 
\end{equation}
Note that the Yukawa interactions needed for the Majorana neutrino
mass necessarily imply the presence of LNV and LFV couplings $h_{ij}$
and the corresponding LFV operator coefficients $c_{\ell}$, 
leading to ${\cal R} \sim O(1)$. 
Again, non-observation of LFV in the next generation of
experiments would typically push $\Lambda$ into the multi-TeV range,
thus implying a negligible short distance contribution to $0 \nu \beta
\beta$.  As with RPV-SUSY, this conclusion can be evaded by assuming a  
specific flavor structure, namely $y_M$ approximately diagonal 
or a nearly degenerate heavy neutrino spectrum.

In both of these phenomenologically viable models
that incorporate LNV and LFV at low scale ($\sim$ TeV), one finds
${\cal R} \gg 10^{-2}$~\cite{Raidal:1997hq,deGouvea:2000cf,Cirigliano:2004mv}.
It is likely that the basic mechanism at work in 
these illustrative cases is  generic: low scale LNV interactions
($\Delta L = \pm 1$ and/or $\Delta L= \pm 2$), which in general
contribute to $0 \nu \beta \beta$, also generate sizable contributions
to $\mu \to e$ conversion, thus enhancing this process over $\mu \to e
\gamma$.

In conclusion, the above considerations suggest that the ratio ${\cal R} = B_{\mu \to
e}/B_{\mu \rightarrow e \gamma}$ of muon LFV processes will provide
important insight about the mechanism of neutrinoless double beta
decay and the use of this process to determine the absolute scale of
neutrino mass.  Assuming observation of LFV processes in forthcoming
experiments, if ${\cal R} \sim 10^{-2}$ the mechanism of $0 \nu
\beta \beta$ is light Majorana neutrino exchange;  if ${\cal R}
\gg 10^{-2}$, there might be TeV scale LNV dynamics, and no definite
conclusion on the mechanism of $0 \nu \beta \beta$ can be drawn based
only on LFV processes.

\section{Overview of the experimental status of search for $\beta\beta$ decay}

The field has a venerable history. The rate of the $2\nu\beta\beta$ decay was
first estimated by Maria Goeppert-Mayer already in 1937 in her thesis work
suggested by E. Wigner, basically correctly. 
Yet, first experimental observation in a laboratory
experiment was achieved only in 1987, fifty years later.
Why it took so long?  As pointed out above, the typical half-life
of the $2\nu\beta\beta$ decay is $\sim 10^{20}$ years. Yet, its 
``signature" is very similar to natural radioactivity, present to some extent
everywhere, and governed by the half-life of $\sim 10^{10}$ years.
So, background suppression is the main problem to overcome
when one wants to study either of the $\beta\beta$ decay modes.

During the last two decades the $2\nu\beta\beta$ decay has been observed
in ``live" laboratory experiments
in many nuclei, often by different groups and using different 
methods. That shows not only the ingenuity of the experimentalists who
were able to overcome the background nemesis, but makes it possible
at the same time to extract the corresponding $2\nu$ nuclear matrix element
from the measured decay rate. In the $2\nu$ mode the half-life is given by
\begin{equation}
1/T_{1/2} = G^{2\nu}(Q,Z) |M^{2\nu}|^2 ~,
\end{equation}
where $G^{2\nu}(Q,Z) $ is an easily and accurately calculable phase space factor.

\begin{figure}
\centerline{\psfig{file=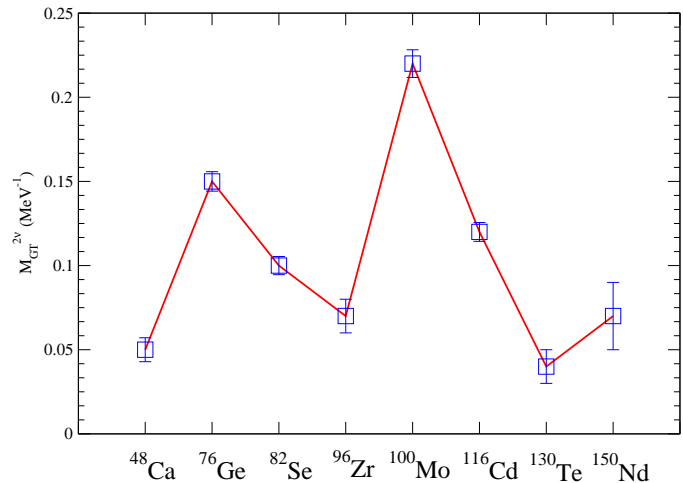,width=8.0cm}}
\caption{Nuclear matrix elements for the $2\nu\beta\beta$ decay extracted from the
measured half-lives.}
\label{fig_m2nu}
\end{figure}

The resulting nuclear matrix elements $M^{2\nu}$, which have the dimension energy$^{-1}$,
are plotted in Fig.\ref{fig_m2nu}. Note the pronounced shell dependence; the matrix element
for $^{100}$Mo is almost ten times larger than the one for $^{130}$Te. Evaluation of these
matrix elements, to be discussed below,
 is an important test for the nuclear theory models that aim at the determination
of the analogous but different quantities for the $0\nu$ neutrinoless mode.

The challenge of detecting the $0\nu\beta\beta$ decay is, at first blush, easier. Unlike the
continuous $2\nu\beta\beta$ decay spectrum with a broad
maximum at rather low energy where the background
suppression is harder, the $0\nu\beta\beta$ decay spectrum is 
sharply peaked at the known $Q$ value (see Fig.\ref{fig_2nu}),
at energies that are not immune to the background, but a bit less difficult to manage.
However, as also indicated in Fig.\ref{fig_2nu}, to obtain interesting results at the
present time means to reach sensitivity to the $0\nu$ half-lives that are $\sim10^6$ times
longer than the $2\nu$ decay half-life of the same nucleus.
So the requirements of background suppression are correspondingly even more severe.

\begin{figure}
\centerline{\psfig{file=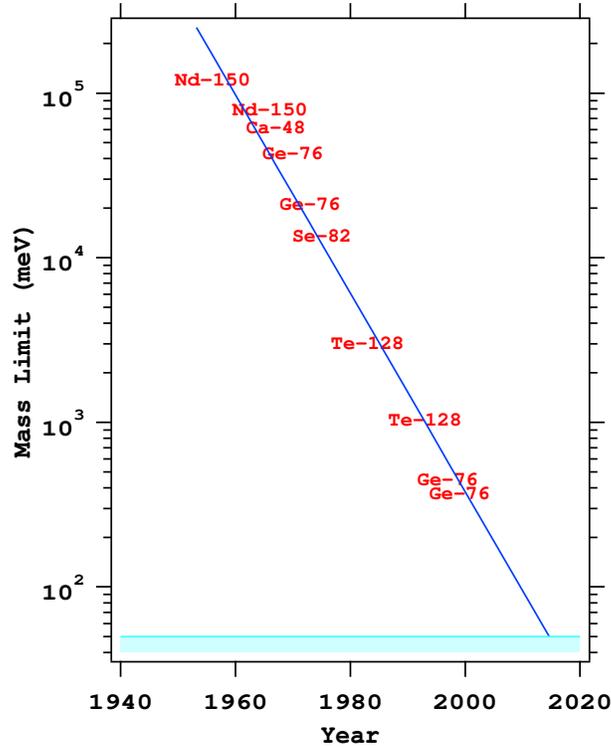,width=9.0cm}}
\caption{The limit of the effective mass $\langle m_{\beta \beta} \rangle$ 
extracted from the experimental lower limits on the $0\nu\beta\beta$ decay half-life
versus the corresponding year. The gray
band near bottom indicates the $\sqrt{\Delta m^2_{atm}}$
value. Figure originally made by S. Elliott.}
\label{fig_history}
\end{figure}

The historical lessons are illustrated in Fig.\ref{fig_history} where the past limits on the
$0\nu\beta\beta$ decay half-lives of various candidate nuclei are translated using 
the eq.(\ref{eq_rate})   into the limits on the effective mass $\langle m_{\beta \beta} \rangle$.
When plotted in the semi-log plot this figure represents the ``Moore's law" of double beta decay,
and indicates that, provided that the past trend will continue, the mass scale corresponding to 
$\Delta m^2_{atm}$ will be reached in about 10 years. 
This is also the time scale of significant experiments these days. Indeed, as discussed further,
preparations are on the way to reach this sensitivity goal.
Note that the figure was made using
some assumed values of the corresponding nuclear matrix elements, without including
their uncertainty. For such illustrative purposes they are, naturally, irrelevant.  

The past search for the neutrinoless double beta decay, illustrated in Fig.\ref{fig_history},
was driven by the then current technology and the resources of the individual experiments.
The goal has been simply to reach sensitivity to longer and longer half-lives.
The situation is different, however, now. The experimentalists at the 
present time can and do use the knowledge
summarized in Fig.\ref{fig_5} to gauge the aim of their proposals. Based on that figure,
the range of the mass parameter $\langle m_{\beta \beta} \rangle$  can be divided
into three regions of interest.
\begin{itemize}
\item The degenerate mass region where all $m_i \gg \sqrt{\Delta m^2_{atm}}$. In that
region $\langle m_{\beta \beta} \rangle \ge$ 0.1 eV, corresponding crudely to the $0\nu$
half-lives of $10^{26-27}$ years. To explore it (in a realistic time frame), 
$\sim$ 100 kg of the decaying nucleus is needed. Several experiments aiming at such 
sensitivity are being built and should run and give results within the next 3-5 years.
Moreover, this mass region (or a substantial part of it) will be explored, in a similar
time frame, by the study of ordinary $\beta$ decay (in particular of tritium) and by
the observational cosmology. These techniques are independent on the Majorana
nature of neutrinos. It is easy, but perhaps premature, 
to envision various scenarios depending on the possible
outcome of these measurements.
\item The so-called inverted hierarchy mass region where 
$20 < \langle m_{\beta \beta} \rangle < 100$ meV and the $0\nu\beta\beta$ half-lives
are about $10^{27-28}$ years. (The name is to some extent a misnomer. In that interval
one could encounter not only the inverted hierarchy but also a quasi-degenerate but
normal neutrino mass ordering. Successful observation of the $0\nu\beta\beta$ decay
will not be able to distinguish these possibilities, as I argued above.
This is so not only due to the anticipated experimental accuracy, but more fundamentally
due to the unknown Majorana phases.) To explore this
mass region, $\sim$ ton size sources would be required. Proposals for the corresponding
experiments exist, but none has been funded as yet, and presumably the real work will
begin depending on the experience with the various $\sim$ 100 kg size sources. Timeline
for exploring this mass region is $\sim$ 10 years.
\item Normal mass hierarchy region where $\langle m_{\beta \beta} \rangle \le$ 10-20 meV.
To explore this mass region, $\sim$ 100 ton sources would be required. There are no realistic
proposals for experiments of this size at present.
\end{itemize}

\begin{figure}
\centerline{\psfig{file=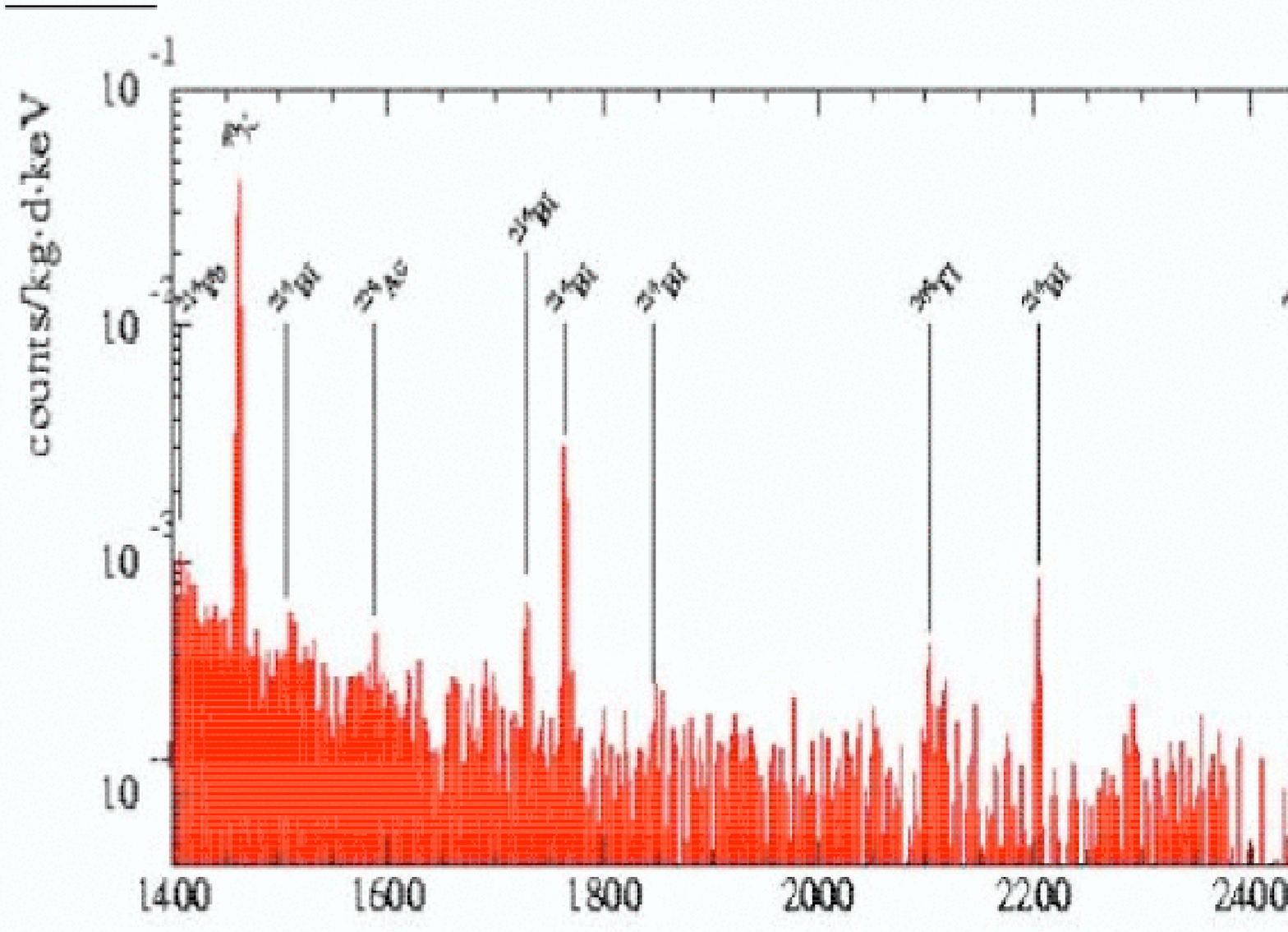,width=12.0cm}}
\caption{The spectrum recorded in the Heidelberg-Moscow $\beta\beta$ decay experiment on
$^{76}$Ge. Identified $\gamma$ lines are indicated.}
\label{fig_HM}
\end{figure}

\begin{figure}
\centerline{\psfig{file=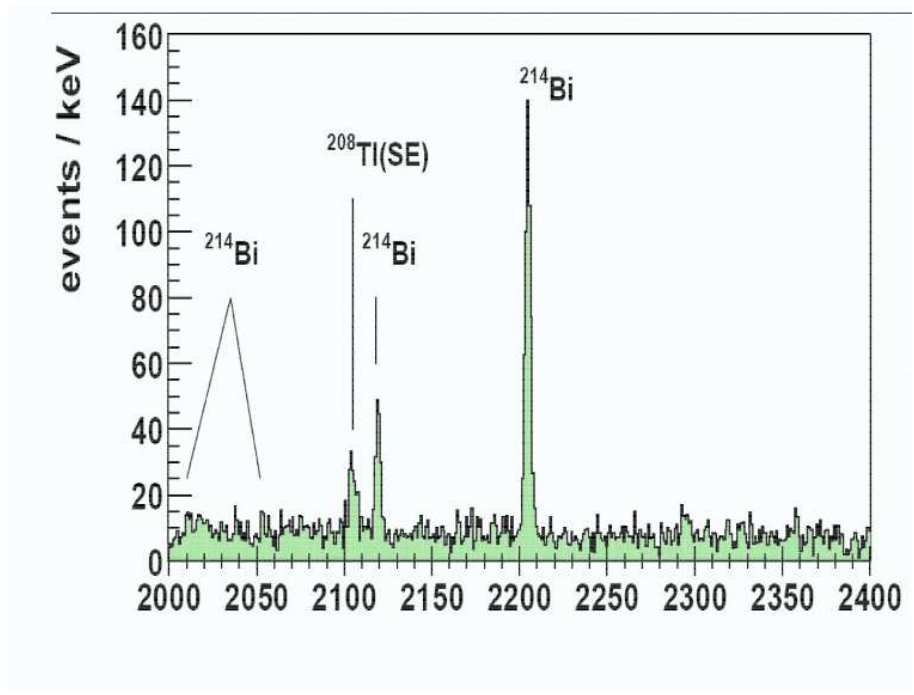,width=12.0cm}}
\caption{Spectrum of the Heidelberg-Moscow experiment in the vicinity of the $0\nu\beta\beta$
decay value of 2039 keV. }
\label{fig_HM_detail}
\end{figure}

Over the last two decades, the methodology for double beta decay experiments has improved
considerably.
Larger amounts of high-purity enriched parent isotopes, combined with careful selection
of all surrounding materials and using deep-underground sites have lowered backgrounds 
and increased sensitivity. The most sensitive experiments to date use $^{76}$Ge, $^{100}$Mo,
$^{116}$Cd, $^{130}$Te, and $^{136}$Xe. For $^{76}$Ge the lifetime limit reached impressive 
values exceeding $10^{25}$years\cite{HM,IGEX}. The experimental lifetime limits have
been interpreted to yield effective neutrino mass limits typically a few eV and in $^{76}$Ge
as low as 0.3 - 1.0 eV (the spread reflects an 
estimate of the uncertainty in the nuclear matrix elements).
The sum electron spectrum obtained in the Heidelberg-Moscow\cite{HM} experiment is shown
in Fig.\ref{fig_HM} over a broad energy range, and in Fig.\ref{fig_HM_detail} over a narrower
range in the vicinity of the $0\nu$ $Q$ value of 2039 keV.  Some residual natural radioactivity
background lines are clearly visible in both figures, and no obvious peak at the $0\nu$ expected
position can be seen in Fig.\ref{fig_HM_detail}.

Nevertheless, a subset of members of the Heidelberg-Moscow collaboration reanalyzed the
data (and used additional information, e.g. the pulse-shape analysis and a different
algorithm in the peak search) and claimed to observe
a positive signal corresponding to the effective mass of     
$\langle m_{\beta \beta} \rangle = 0.39_{-0.28}^{+0.17}$ eV\cite{Klapdor}. 
That report has been followed by a lively discussion.  Clearly, such an extraordinary claim
with its profound implications, requires extraordinary evidence. It is fair to say that
a confirmation, both for the same $^{76}$Ge parent nucleus, and better yet also in another
nucleus with a different $Q$ value, would be required for a consensus. In any case, if that
claim is eventually confirmed, the degenerate mass scenario will be implicated, and eventual
positive signal in the analysis of the tritium $\beta$ decay and/or observational cosmology
should be forthcoming. For the neutrinoless $\beta\beta$ decay the next generation of experiments,
which  use $\sim$ 100 kg of decaying isotopes will, among other things, test this
recent claim. 

It is beyond the scope of these lecture notes to describe 
in detail the forthcoming $0\nu\beta\beta$
decay experiments. Rather detailed discussion of them can be found e.g. in Ref.\cite{EE04}.
Also, the corresponding chapter of the APS neutrino study\cite{APS} has various details.
Nevertheless, let me briefly comment on the most advanced of the forthcoming $\sim$ 100 kg
source experiments {\it CUORE, GERDA, EXO, {\rm and} MAJORANA}. 
All of them are designed to explore
all (or at least most) of the degenerate neutrino mass region 
$\langle m_{\beta \beta} \rangle \ge$
0.1 eV. If their projected efficiencies and background projections are confirmed, all of them
plan to consider scaling up the decaying mass to $\sim$ ton and extend their sensitivity to the
``inverted hierarchy" region.

These experiments use different nuclei as a source, $^{76}$Ge for {\it GERDA} and {\it MAJORANA},
$^{130}$Te for {\it CUORE}, {\rm and} $^{136}$Xe for {\it EXO}. 
The requirement of radiopurity of the source
material and surrounding auxiliary equipment is common to all of them, as is the placement 
of the experiment deep underground to shield against cosmic rays. The way the electrons
are detected is, however, different. While the germanium detectors with their superb energy
resolution have been used for the search of the $0\nu\beta\beta$ decay for a long time,
the cryogenic detectors in {\it CUORE} use the temperature increase associated with an event
in the very cold TeO$_2$ crystals, and in the {\it EXO} experiment a Time Projection Chamber (TPC)
uses both scintillation and ionization to detect the events. The {\it EXO} experiment in its final form
(still under development and very challenging) would use a positive identification of the final
Ba$^+$ ion as an ultimate background rejection tool. These four experiments are in various stages
of funding and staging. First results are expected in about 3 years, and substantial results within
3-5 years in all of them.

\section{Nuclear matrix elements}

It follows from eq.(\ref{eq_rate}) that (i) values of the nuclear matrix elements
$M^{0\nu}$ are needed in order to extract the effective neutrino mass from
the measured $0\nu\beta\beta$ decay rate, and (ii) any uncertainty in  $M^{0\nu}$ 
causes a corresponding and equally large uncertainty in the 
extracted $\langle m_{\beta \beta} \rangle$ value. Thus, the issue of an accurate
evaluation of the nuclear matrix elements attracts considerable attention.

To see qualitatively where the problems are, let us consider the so-called closure
approximation, i.e. a description in which the second order perturbation expression 
is approximated as
\begin{equation}
M^{0\nu} \equiv \langle \Psi_{final} | \hat{O}^{(0\nu)} | \Psi_{initial} \rangle ~.
\end{equation}

Now, the challenge is to use an appropriate many-body nuclear model to describe
accurately the wave functions of the ground states of the initial and final nuclei,
$|  \Psi_{initial} \rangle$ and  $|  \Psi_{final} \rangle$, as well as the appropriate
form of the effective transition  operator $\hat{O}^{(0\nu)}$ that describes the
transformation of two neutrons into two protons correlated by the neutrino propagator,
and consistent with the approximations inherent to the nuclear model used.

Common to all methods is the description of the nucleus as a system of nucleons bound
in the mean field and interacting by an effective residual interaction. The used methods
differ as to the number of nucleon orbits (or shells and subshells) included in 
the calculations and
the complexity of the configurations of the nucleons in these orbits. 
The two basic approaches used so far
for the evaluation of the nuclear matrix elements for both the $2\nu$ and $0\nu$ 
$\beta\beta$ decay modes
are the Quasiparticle Random Phase Approximation (QRPA) and the nuclear shell model (NSM).
They are in some sense complementary; QRPA uses a larger set of orbits, but truncates
heavily the included configurations, while NSM can include only a rather small set
of orbits but includes essentially all possible configurations. NSM also can be tested
in a considerable detail 
by comparing to the nuclear spectroscopy data; in QRPA such comparisons are much
more limited.

 For the $2\nu$ decay one can relate the various factors entering the calculations to other
 observables ($\beta$ strength functions, cross sections of the charge-exchange
 reactions, etc.), accessible to the experiment. The consistency of the evaluation can be
 tested in that way. Of course, as pointed out above (see Fig.\ref{fig_m2nu}) the nuclear
 matrix elements for this mode are known anyway. Both methods are capable of describing
the $2\nu$ matrix elements, at least qualitatively. These quantities, when expressed in
natural units based on the sum rules, are very small. Hence their description depends 
on small components of the nuclear wave functions and is therefore challenging.
In QRPA the agreement is achieved if the effective proton-neutron interaction
coupling constant (usually called $g_{pp}$) is slightly (by $\sim$ 10 - 20 \%)
adjusted.
 
 The theoretical description for the more interesting $0\nu$ mode cannot use
 any known nuclear observables, since there are no observables
 directly related to the $M^{0\nu}$. It is therefore much less clear how to properly
 estimate the uncertainty associated with the calculated values of  $M^{0\nu}$,
 and to judge their accuracy.
 Since the calculations using QRPA are much simpler, an overwhelming majority
 of the published calculations uses that method. There are suggestions to use the
 spread of these published values of   $M^{0\nu}$ as a measure of uncertainty\cite{BM04}.
 Following this, one would conclude that the uncertainty is quite large, a factor of three
 or as much as five. But that way of assigning the uncertainty is questionable. Using all
 or most of the published values of $M^{0\nu}$ means that one includes calculations
 of uneven quality. Some of them were devoted to the tests of various approximations,
 and concluded that they are not applicable. Some insist that other data, like the
 $M^{2\nu}$, are correctly reproduced, other do not pay any attention to such test.
 Also, different forms of the transition operator $\hat{O}^{0\nu}$ are used, in particular
 some works include approximately the effect of the short  range nucleon-nucleon repulsion,
 while others neglect it. 

\begin{figure}
\centerline{\psfig{file= 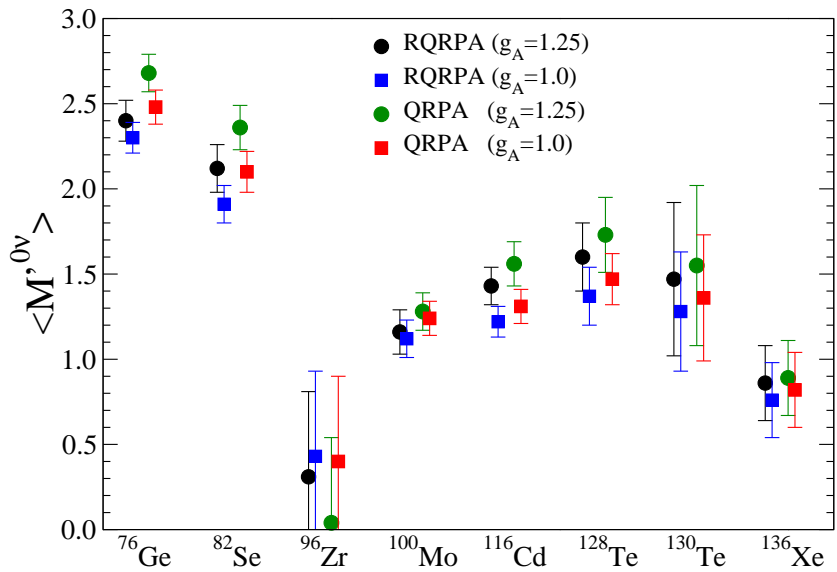,width=11.0cm}}
\caption{Nuclear matrix elements and their variance for the indicated approximations
(see Ref.\cite{Rodin}). }
\label{fig_nuclme}
\end{figure}

In contrast, in Ref.\cite{Rodin} an assesment of uncertainties
in the matrix elements $M^{0\nu}$ inherent in the QRPA was made, and it was concluded that 
with a consistent treatment the uncertainties are much less, perhaps only about 30\%
(see Fig.\ref{fig_nuclme}).
That calculation uses the known $2\nu$ matrix elements in order to adjust the
interaction constant mentioned above.
There is a lively debate in the nuclear structure theory community, 
beyond the scope of these lectures, about this conclusion. 

\vspace{1cm}

It is of interest also to compare the resulting matrix elements 
of Rodin {et al.}\cite{Rodin} based
on QRPA and its generalizations, 
and those of the available most recent NSM evaluation\cite{Poves}. 
Note that the operators used in NSM evaluation do not include the induced nucleon currents
that in QRPA reduce the matrix element by about 30\%. The QRPA\cite{Rodin}
and NSM\cite{Poves} $M^{0\nu}$ are compared in Table \ref{tab_nme}. In the last column 
the NSM matrix elements are reduced by 30\% to approximately account for the missing terms
in the operator, and to make the comparison more meaningful. With this reduction,
it seems that QRPA results are a bit larger in the lighter nuclei and a bit smaller in the heavier
ones than the NSM results, but basically within the 30\% uncertainty estimate. 
Once the NSM calculations for the intermediate mass nuclei $^{96}$Zr, $^{100}$Mo
and $^{116}$Cd become available, one can make a more meaningful comparison of the two
methods.

\begin{table}
\begin{center}
\tbl{Comparison of the calculated nuclear matrix elements $M^{0\nu}$ using the QRPA
method\cite{Rodin} and the NSM\cite{Poves}. In the last column the NSM values are reduced,
divided by 1.3, to account approximately for the effects of the induced nucleon currents. } 
{\begin{tabular}{|c|c|c|c|} \toprule
Nucleus & QRPA & NSM & NSM/1.3 \\
\hline
$^{76}$Ge & 2.3-2.4 & 2.35 & 1.80 \\
$^{82}$Se & 1.9-2.1 & 2.26 & 1.74 \\
$^{96}$Zr & 0.3-0.4  &           &          \\
$^{100}$Mo & 1.1-1.2 &         &         \\
$^{116}$Cd &  1.2-1.4 &        &         \\
$^{130}$Te  &   1.3     &  2.13 &   1.64 \\
$^{136}$Xe  &    0.6-1.0 & 1.77 & 1.36 \\
\hline
\label{tab_nme}
\end{tabular}}
\end{center}
\end{table}

When comparing the results shown in Table \ref{tab_nme} as well as the results
of other calculations (e.g. Refs.\cite{Suh1,Suh2} ) with Fig. \ref{fig_m2nu} it is important
to notice a qualitative difference in the behaviour of the $2\nu$ and $0\nu$ matrix elements
when going from one nucleus to another one. For $2\nu$ the matrix elements change 
rapidly, but for the $0\nu$ the variation is much more gentle ($^{96}$Zr is a notable exception,
at least for QRPA). That feature, common to most calculations, if verified,  
would help tremendously in comparing
the results or constraints from one nucleus to another one.

Once the nuclear matrix elements are fixed (by choosing your favorite set of results),
they can be combined with the phase space factors (a complete list is available, e.g.
in the monograph\cite{BV92}) to obtain a half-life prediction for any value of the effective
mass   $\langle m_{\beta \beta} \rangle$. 
It turns out that for a fixed $\langle m_{\beta \beta} \rangle$
the half-lives of different candidate nuclei do not differ very much 
from each other (not more than by factors $\sim 3$ or so)
and, for example, the boundary between the degenerate and inverted hierarchy mass
regions corresponds to half-lives $\sim10^{27}$years. Thus, the next generation
of experiments, discussed above, should reach this region using several candidate
nuclei, making the corresponding conclusions less nuclear model dependent. 

\section{Neutrino magnetic moment and the distinction between Dirac and Majorana
neutrinos}

Neutrino mass and magnetic moments are intimately related. In the orthodox Standard
Model neutrinos have a vanishing mass and  magnetic moments vanish as well. 
However, in the 
minimally extended SM containing gauge-singlet right-handed neutrinos the
magnetic moment $\mu_{\nu}$ is nonvanishing
and proportional to the neutrino mass, but unobservably small \cite{musm},
\begin{equation}
\mu_{\nu} = \frac{3eG_F}{\sqrt{2}8\pi^2} m_{\nu} 
=  3 \times 10^{-19} \mu_B \frac{m_{\nu}}{1{~\rm eV}} ~.
\label{eq_munu}
\end{equation}
Here $\mu_B$ is the electron Bohr magneton, traditionally used as unit also
for the neutrino magnetic moments.
An experimental observation of a magnetic moment larger than that given in
eq.(\ref{eq_munu}) would be an uneqivocal indication of physics beyond the
minimally extended Standard Model.

Laboratory searches for neutrino magnetic moments are typically based on the obsevation
of the $\nu - e$ scattering. Nonvanishing $\mu_{\nu}$ will be recognizable only if the
corresponding electromagnetic scattering cross section is at least comparable to the well
understood weak interaction cross section. The magnitude of $\mu_{\nu}$ 
(diagonal in flavor or transitional) which can 
be probed in this way  is then given by
\begin{equation}
\frac {|\mu_{\nu}|}{\mu_B} \equiv \frac{G_F m_e}{\sqrt{2} \pi \alpha} \sqrt{m_e T} \sim 10^{-10}
\left(\frac{T}{m_e} \right)^{1/2} ~,
\label{eq_muesc}
\end{equation}
where $T$ is the electron recoil kinetic energy. Considering realistic values of $T$, it would
be difficult to reach sensitivities below $\sim 10^{-11} \mu_B$. Present limits are about
an order of magnitude larger than that.

Limits on $\mu_{\nu}$ can also be obtained from bounds on the unobserved energy loss in
astrophysical objects. For sufficiently large $\mu_{\nu}$ the rate of plasmon decay 
into the $\nu \bar{\nu}$ pairs  would conflict with such bounds. Since plasmons can also decay
weakly into the $\nu \bar{\nu}$ pairs , the sensitivity of this probe is again limited by the size
of the weak rate, leading to
\begin{equation}
\frac {|\mu_{\nu}|}{\mu_B} \equiv \frac{G_F m_e}{\sqrt{2} \pi \alpha}{\hbar \omega_P} ~,
\end{equation}
where $\omega_P$ is the plasmon frequency. Since usually $(\hbar \omega_P)^2 \ll m_e T$
that limit is stronger than that given in eq.(\ref{eq_muesc}). Current limits on $\mu_{\nu}$ 
based on such considerations are $\sim 10^{-12} \mu_B$.

The interest in $\mu_{\nu}$ and its relation to neutrino mass dates from $\sim$1990
when it was suggested that the chlorine data\cite{chlorine} on solar neutrinos show
an anticorrelation between the neutrino flux and the solar activity characterized by the
number of sunspots. A possible explanation was suggested in Ref.\cite{Okun} where
it was proposed that a magnetic moment $\mu_{\nu} \sim 10^{-(10-11)} \mu_B$ would
cause a precession in solar magnetic field of the neutrinos emitted initially as 
left-handed $\nu_e$ into unobservable right-handed ones. Even though
later analyses showed that the 
correlation with solar acivity does not exist, the possibility of a relatively large
$\mu_{\nu}$ accompanied by a small mass $m_{\nu}$ was widely discussed and various models 
accomplishing that were suggested.

 If a magnetic moment is generated by physics
beyond the Standard Model (SM) at an energy scale $\Lambda$, 
we can generically express its value as
\begin{equation}
\mu_\nu \sim \frac{eG}{\Lambda},
\end{equation}
where $e$ is the electric charge and $G$ contains a combination of
coupling constants and loop factors.  Removing the photon from the 
diagram  gives a contribution to the neutrino mass of order
\begin{equation}
m_\nu \sim G \Lambda.
\end{equation}
We thus arrive at the relationship
\begin{equation}
m_\nu \sim  \frac{\Lambda^2}{2 m_e}  \frac{\mu_\nu}{\mu_B}
~~ \sim \frac{\mu_\nu}{ 10^{-18} \mu_B}
[\Lambda({\rm TeV})]^2  \,\,\,{\rm eV},
\label{naive}
\end{equation}
which implies that it is difficult to simultaneously reconcile a small
neutrino mass and a large magnetic moment.

 This na\"ive restriction given in
eq.(\ref{naive}) can be overcome via a careful choice for the new
physics, e.g., by requiring certain additional symmetries
\cite{Vol88,Bar89,Geo90,Gri91,Bab90,Barr90}. Note,
however, that these symmetries are typically broken by Standard Model
interactions.
For Dirac neutrinos such symmetry (under which the left-handed neutrino 
and antineutrino $\nu$ and $\nu^c$ transform as a doublet)
is violated by SM gauge interactions. For Majorana neutrinos
analogous symmetries are not
broken by SM gauge interactions, but are instead violated by SM Yukawa
interactions, provided that the charged lepton masses are generated
via the standard mechanism through Yukawa couplings to the SM Higgs
boson. This suggests that the relation between $\mu_{\nu}$ and $m_{\nu}$
is different for Dirac and Majorana neutrinos. This distinction can be,
at least in principle, exploited experimentally, as shown below.

Earlier, I have quoted the Ref.\cite{SV82} (see Fig.\ref{fig_SV}) to stress that observation
of the $0\nu\beta\beta$ decay would necessarily imply the existence of a
novanishing neutrino Majorana mass. Analogous considerations can be
applied in this case.  By calculating neutrino magnetic moment
contributions to $m_\nu$ generated by SM radiative corrections, one may
obtain in this way general, \lq\lq naturalness" upper limits on the size of
neutrino magnetic moments by exploiting the experimental upper
limits on the neutrino mass.

In the case of Dirac neutrinos, a magnetic moment term will
generically induce a radiative correction to the neutrino mass of
order\cite{mu_D}
\begin{equation}
m_\nu \sim \frac{\alpha}{16\pi}
\frac{\Lambda^2}{m_e}  \frac{\mu_\nu}{\mu_B}
~~ \sim \frac{\mu_\nu}{3 \times 10^{-15} \mu_B}
[\Lambda({\rm TeV})]^2 \,\,\,{\rm eV}.
\end{equation}
Taking $\Lambda \simeq$ 1 TeV and $m_\nu \le$ 0.3 eV, we obtain
the limit $\mu_\nu \le 10^{-15} \mu_B$ (and a more stringent one
for larger $\Lambda)$, which is several orders of
magnitude more constraining than current experimental 
upper limits on $\mu_{\nu}$.

 The case of Majorana neutrinos is more subtle, due to the relative
flavor symmetries of $m_\nu$ and $\mu_\nu$ respectively. 
For Majorana neutrinos  the
transition magnetic moments $\left[\mu_\nu\right]_{\alpha\beta}$
(the only possible ones) are
antisymmetric in the flavor indices $\{\alpha,\beta\}$, while the mass
terms $[m_\nu]_{\alpha\beta}$ are symmetric.  These different flavor
symmetries play an important role in the limits, and are the origin of
the difference between the magnetic moment constraints for Dirac and Majorana
neutrinos.  

It has been shown in Ref.\cite{mu_M} that the constraints on Majorana
neutrinos are significantly weaker than those 
for Dirac neutrinos\cite{mu_D}, as the different flavor symmetries of $m_\nu$ and
$\mu_\nu$ lead to a mass term which is suppressed only by charged lepton
masses.  This conclusion was reached by considering one-loop mixing of the
magnetic moment and mass operators generated by Standard Model interactions.
The authors of Ref.\cite{mu_M} found that
if a magnetic moment arises through a coupling of the
neutrinos to the neutral component of the $SU(2)_L$ gauge boson, the
constraints for $\mu_{\tau e}$ and
$\mu_{\tau\mu}$ are comparable to present experiment limits, while the
constraint on $\mu_{e\mu}$ is significantly weaker.  
Thus, the analysis of  Ref.\cite{mu_M} 
lead to a bound for the transition magnetic moment
of Majorana neutrinos that is less stringent than
present experimental limits.

Even more generally it was shown in Ref.\cite{mu_last}  
that two-loop matching of mass and magnetic moment operators implies 
stronger constraints than those obtained
in\cite{mu_M} if the scale of the new physics $\Lambda \ge 10$
TeV. Moreover, these constraints apply to a magnetic moment generated
by either the hypercharge or $SU(2)_L$ gauge boson.
In arriving at these conclusions, the most general set of operators that
contribute at lowest order to the mass and magnetic moments of
Majorana neutrinos was constructed, and model independent constraints which
link the two were obtained.   Thus the results of Ref.\cite{mu_last} imply
completely model independent naturalness bound that -- for $\Lambda \ge 100$
TeV -- is stronger than present experimental limits (even for the
weakest constrained element $\mu_{e\mu}$). On the other hand, for sufficiently
low values of the scale $\Lambda$ the known small values of the neutrino
masses do not constrain the magnitude of the transition
magnetic moment $\mu_{\nu}$ for Majorana neutrinos
more than the present experimental limits. Thus, if these conditions are fulfilled,
the discovery of $\mu_{\nu}$ might be forthcoming any day.
 
The above result means that an experimental discovery of a magnetic moment
near the present limits would signify that (i) neutrinos are Majorana
fermions and (ii) new lepton number violating physics responsible for
the generation of $\mu_\nu$ arises at a scale $\Lambda$ which is well
below the see-saw scale. This would have,
among other things, implications for the mechanism of the
neutrinoless double beta
decay and lepton flavor violation as discussed above and in Ref.\cite{LNVus}.

\section{Summary}

In these lectures I discussed the status of double beta decay, its relation to
the charge conjugation symmetry of neutrinos and to the problem
of the lepton number conservation in general. I have shown that if one makes
the minimum assumption that the light neutrinos familiar from the oscillation 
experiments which are interacting by the left-handed weak current are Majorana
particles, then the rate of the $0\nu\beta\beta$ decay can be related to the
absolute scale of the neutrino mass in a straightforward way.

On the other hand, it is also possible that the $0\nu\beta\beta$ decay is mediated
by the exchange of heavy particles. I explained that if  the corresponding mass
scale of such hypothetical particles is $\sim$ 1 TeV, the corresponding $0\nu$
decay rate could be comparable to the decay rate associated with the exchange
of a light neutrino. I further argued that the study of the lepton flavor violation
involving $\mu \to e$ conversion and $\mu \to e + \gamma$ decay may be used
as a ``diagnostic tool" that could help to decide which of the possible mechanisms
of the $0\nu$ decay is dominant. 

Further, I have shown that the the range of the effective masses $\langle m_{\beta\beta} \rangle$
can be roughly divided into three regions of interest, each corresponding to a different
neutrino mass pattern. The region of   $\langle m_{\beta\beta} \rangle \ge$ 0.1 eV corresponds
to the degenerate mass pattern. Its exploration is well advanced, and one can rather
confidently expect that it will be explored by several $\beta\beta$ decay experiments in
the next 3-5 years. This region of neutrino masses (or most of it)
is also accessible to studies using the ordinary $\beta$ decay and/or the observational cosmology.
Thus, if the nature is kind enough to choose this mass pattern, we will have a multiple ways
of exploring it.

The region of $0.01 \le \langle m_{\beta\beta} \rangle \le 0.1$ eV 
is often called the "inverted mass
hierarchy" region. In fact, both the inverted and the quasi-degenerate but normal mass orderings
are possible in this case, and experimentally indistinguishable. 
Realistic plans to explore this region
using the $0\nu\beta\beta$ decay exist, but correspond to a longer time scale of about 10 years.
They require much larger, $\sim$ ton size $\beta\beta$ sources and correspondingly even 
more stringent background suppression. 

Finally, the region $\langle m_{\beta\beta} \rangle \le 0.01$ eV 
corresponds to the normal hierarchy
only. There are no realistic proposals at present to explore this mass region
experimentally.

Intimately related to the extraction of $\langle m_{\beta\beta} \rangle$ from the decay rates is
the problem of nuclear matrix elements. At present, there is no consensus among the nuclear
theorists about their correct values, and the corresponding uncertainty. I argued that the 
uncertainty is less than some suggest, and that the closeness of the Quasiparticle Random
Phase Approximation (QRPA) and Shell Model (NSM) results are encouraging. But this
is still a problem that requires further improvements.

In the last part 
I discussed the neutrino magnetic moments. I have shown that using the Standard Model
radiative correction one can calculate the contribution of the magnetic moment to the neutrino
mass. That contribution, naturally, should not exceed the experimental upper limit on the
neutrino mass. Using this procedure one can show that the magnetic moment of Dirac neutrinos
cannot exceed about $10^{-15} \mu_B$, which is several orders of magnitudes less than the
current experimental limits on $\mu_{\nu}$. On the other hand, due to the different symmetries of the
magnetic moment and mass matrices for Majorana neutrinos, the corresponding constraints
are much less restrictive, and do not exceed the current limits. Thus, a discovery of $\mu_{\nu}$
near the present experimental limit would indicate that neutrinos are Majorana particles,
and the corresponding new physics scale is well below the GUT scale. 

\section{Acknowlegment}

The original results reported here were obtained in the joint 
and enjoyable work
with a number of collaborators, Nicole Bell, Vincenzo Cirigliano, Steve Elliott,
Amand Faessler, Michail Gorchtein, Andriy Kurylov, Gary Prezeau, Michael Ramsey-Musolf, 
Vadim Rodin, Fedor \v{S}imkovic and Mark Wise.
The work was supported in part under U.S. DOE contract DE-FG02-05ER41361.

\bibliographystyle{ws-rv-van}
\bibliography{ws-rv-sample}

\end{document}